\documentclass[sigconf,pbalance]{acmart}
\usepackage{booktabs}
\usepackage{graphicx}

\setcopyright{acmlicensed}
\copyrightyear{2027}
\acmYear{2027}
\begin{document}

\setcopyright{none}

\title[Weight Coding: Evolving Web-Scale Recommenders]{Weight Coding: Evolving Web-Scale Recommenders through Direct Weight Edits with LLM Agents}

\author{Junpeng Hou}
\authornote{Corresponding author ryanhou94@gmail.com.}
\email{jhou@pinterest.com}
\affiliation{%
  \institution{Pinterest, Inc.}
  \city{San Francisco}
  \state{CA}
  \country{USA}
}

\author{Jiaxing Qu}
\email{jqu@pinterest.com}
\affiliation{%
  \institution{Pinterest, Inc.}
  \city{San Francisco}
  \state{CA}
  \country{USA}
}

\author{Derek Cheng}
\email{naichiacheng@pinterest.com}
\affiliation{%
  \institution{Pinterest, Inc.}
  \city{San Francisco}
  \state{CA}
  \country{USA}
}

\author{Jaewon Yang}
\email{jaewonyang@pinterest.com}
\affiliation{%
  \institution{Pinterest, Inc.}
  \city{San Francisco}
  \state{CA}
  \country{USA}
}

\author{Abhinav Naikawadi}
\email{anaikawadi@pinterest.com}
\affiliation{%
  \institution{Pinterest, Inc.}
  \city{San Francisco}
  \state{CA}
  \country{USA}
}

\author{Karthik Subbian}
\email{ksubbian@pinterest.com}
\affiliation{%
  \institution{Pinterest, Inc.}
  \city{San Francisco}
  \state{CA}
  \country{USA}
}

\begin{CCSXML}
<ccs2012>
<concept><concept_id>10002951.10003317.10003347.10003350</concept_id><concept_desc>Information systems~Recommender systems</concept_desc><concept_significance>500</concept_significance></concept>
<concept><concept_id>10010147.10010257</concept_id><concept_desc>Computing methodologies~Machine learning</concept_desc><concept_significance>500</concept_significance></concept>
<concept><concept_id>10010147.10010178.10010219.10010221</concept_id><concept_desc>Computing methodologies~Intelligent agents</concept_desc><concept_significance>300</concept_significance></concept>
</ccs2012>
\end{CCSXML}
\ccsdesc[500]{Information systems~Recommender systems}
\ccsdesc[500]{Computing methodologies~Machine learning}
\ccsdesc[300]{Computing methodologies~Intelligent agents}
\keywords{Weight coding, Recommender systems, Direct weight editing, LLM agents, Model evolution, Gradient-free optimization}

\begin{abstract}
Web-scale recommender models evolve primarily through retraining or fine-tuning. We propose \emph{weight coding}, a methodology for model evolution through direct, executable weight edits. A large language model (LLM) agent inspects model structure and parameters, formulates hypotheses, and implements edits using evaluation feedback. The loop progresses from localized changes to targeted searches and composition, allowing edited components to adapt to one another. A cached engine enables rapid forward evaluation without backpropagation. On Pinterest's 147M-parameter Image Search two-tower retriever, weight coding improves held-out recall@10 of production model by up to 15.0 percentage points (+24\% relative) with few edits in about 1/4 GPU hours of a full post-training finetune, while a common launch only improves around 1\%. After reinitializing the entire 13.6M-parameter interaction module randomly while retaining the surrounding trained model, searching 2,448 mixing and normalization parameters yields 9.8 points above production. Offline gains span the full funnel including dense retrieval, generative retrieval and ranking, with up to 1.37B learnable parameters and different edits for each architecture. Steering edits provide explicit control of model performance over segments like freshness and product. Weight coding requires no serving change and online A/B tests increase saves per user from the edited retrieval source by 8.4\% to 10.7\%, aligned with the target business objectives. Our results establish weight coding as a practical and useful tool for model development and and behavioral control in a real web-scale production system, with potential to extend to both pre-training and post-training stages across the funnel.
\end{abstract}

\begin{teaserfigure}
\centering
\includegraphics[width=\textwidth]{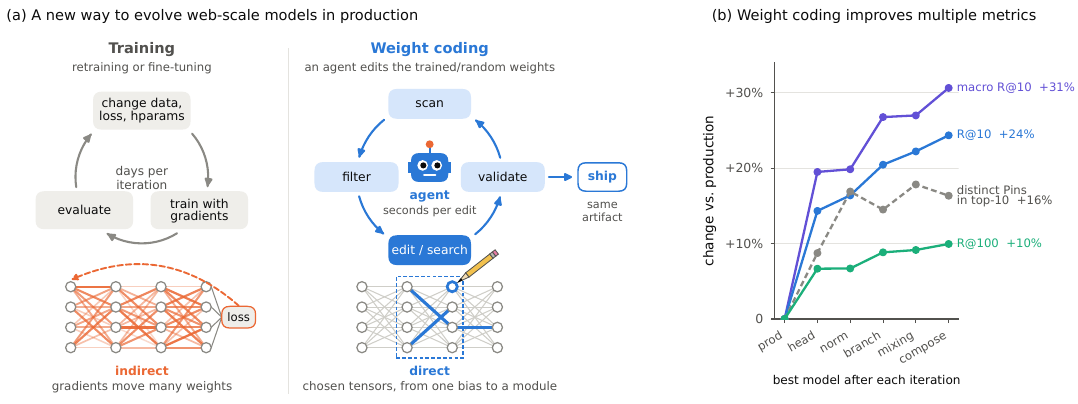}
\caption{Weight coding by LLM agents. (a) Training and direct editing of trained or randomized weights through an agent-driven loop. (b) Relative metric changes after successive weight-coding iterations on the Image Search retrieval model at Pinterest.}
\label{fig:1}
\end{teaserfigure}

\maketitle

\section{Introduction}\label{sec:1}

Web-scale recommender models adapt to changing user interests, content inventory, product requirements and business objectives. Conventional development changes data, losses or architectures and then retrains (pre-training) or fine-tunes (post-training) the model to achieve that \cite{covington2016youtube,yi2019sampling}. The weights themselves offer another point of access. They can be inspected and modified directly, with each change evaluated on the task of interest. \emph{Can models evolve through explicit, executable edits to their weights, guided by feedback from complicated business goals?}

We propose \emph{weight coding}, a methodology for evolving models through direct weight edits (Fig.~\ref{fig:1}(a)). As in software development, changes can be inspected, tested, revised and composed. The process improves task performance and directs model behavior, starting from trained parameter values or reinitialized components.

In weight coding, a large language model (LLM) agent inspects model structure and weights, formulates hypotheses and executes edits guided by evaluation feedback. Given the large scale of production models and evaluators, a cached engine is introduced to support rapid forward evaluation, while held-out dates and production evaluators test the resulting models. Human engineers set the goals and approve deployment to hold accountability and responsibility for online results and business outcomes.

\begin{figure*}[t]
\centering
\includegraphics[width=\textwidth]{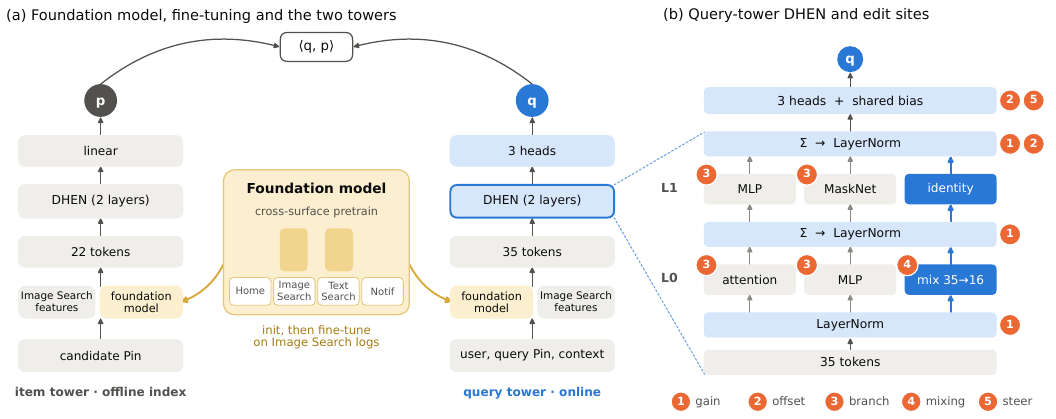}
\caption{Image Search two-tower retrieval model at Pinterest. (a) Both towers start from a cross-surface foundation model and are fine-tuned on Image Search logs. (b) Query-tower DHEN and the edit sites used in this paper.}
\Description{Two panels show foundation-model fine-tuning for a two-tower retriever and the query-tower DHEN architecture with its edit sites.}
\label{fig:2}
\end{figure*}

Our main case is Pinterest's 147M-parameter Image Search retriever \cite{liu2017relatedpins}, which retrieves multi-modal content items called Pins. This two-tower model uses a DHEN interaction module \cite{zhang2022dhen} (Fig.~\ref{fig:2}) and simply resetting certain normalization parameters improves recall@10 by 10.1 percentage points. Combining edits and searches raises the gain to 15.0 points (+24\% relative) on held-out days under the production evaluator. Fig.~\ref{fig:1}(b) previews this progression.

We then ask whether improvements require trained weights. After reinitializing the entire 13.6M-parameter DHEN while retaining the surrounding trained model, searching only 2,448 mixing and normalization parameters yields 9.8 recall@10 points above production. Direct editing thus extends from improving trained parameter values to establishing effective computation within a randomized module, suggesting a broader role for weight coding in both pre- and post-training.

Offline gains extend across the recommendation funnel. On PinRec \cite{botta2025pinrec}, a generative retrieval model with GPT-backbone deployed at Pinterest, bias edits and checkpoint interpolation improve unordered recall@1 by 2.3 points. On the much larger and more complicated ranker, weakening experts improves ranking metrics on every held-out day. The process carries over while the exact edits differ from model to model. Steering also controls model behavior explicitly for different business objectives, adjusting fresh Pins' share in the top-10 from 2\% to 70\% through one coefficient, with a separate direction controlling product exposure.

Online A/B tests of the edited two-tower retriever increase closeups by up to 7.6\% and saves by up to 10.7\% per user for the edited retrieval source, with engagement and exposure changes aligned with the objectives. These results connect direct model evolution to real user behavior and business outcomes. Our contributions are
\begin{itemize}
\item \textbf{A practical methodology for web-scale model development.} We organize executable weight edits into a process run by an LLM agent and guided by task feedback in real production environments.
\item \textbf{Improvement on both trained and randomized weights.} We improve production models from trained weights and construct effective computation within a reinitialized interaction module with random weights.
\item \textbf{Validation across the funnel and online.} We demonstrate offline gains across retrieval and ranking, explicit model behavioral controls for different business objectives and online engagement improvements.
\end{itemize}

\section{Related Work}\label{sec:2}

\textbf{Editing weights and representations.} Knowledge editing and changes to prediction rules target associations or behaviors \cite{meng2022rome,meng2023memit,mitchell2022mend,fang2025alphaedit,santurkar2021editing}, evaluated through efficacy, generalization and specificity \cite{yao2023editing,wang2024survey}. Weight arithmetic combines fine-tuning deltas or snapshots \cite{ilharco2023task,wortsman2022soups,wortsman2022wiseft,yadav2023ties,yu2024dare}. Activation steering changes hidden states along directions \cite{turner2023actadd,zou2023repe,li2023iti,arditi2024refusal}, drawing on linear representations \cite{park2024linear}. Rank reduction \cite{sharma2024laser}, pruning \cite{frankle2019lottery,ramanujan2020hidden,michel2019heads} and training normalization parameters on random features \cite{frankle2021bn} show that useful computation can survive substantial weight changes or arise without optimizing every parameter. Low intrinsic dimensionality offers related evidence \cite{li2018intrinsic,aghajanyan2021intrinsic}. Weight coding organizes such operations into an iterative process for model evolution in real production.

\textbf{Recommender interventions.} Bias correction uses training-time intervention or weighting \cite{zhang2021pda,wei2021macr,schnabel2016treatments} and post hoc score priors \cite{menon2021logit,agarwal2025needle}. PPD removes an embedding popularity direction \cite{islam2026ppd}, while SPREE and PopSteer steer activations \cite{schirmer2026spree,ahmadov2026popsteer}. Recommendation editing removes unwanted recommendations or adds cold-start items \cite{lai2024editing,shen2026genrecedit}, and merging supports cross-domain or multi-objective models \cite{kim2026mergerec,jia2026recsoup,hou2025weaverec}. Industrial controllability uses condition inputs or distribution models \cite{liu2024crm,hou2026causal}. Interaction architectures \cite{zhang2022dhen,wang2021masknet,zhang2024wukong} and embedding collapse \cite{guo2024collapse} identify structure relevant to editing. We examine a common editing process across production retrieval and ranking.

\textbf{Optimization and agents.} Evolution strategies (ES) \cite{salimans2017es,qiu2026es} and zeroth-order methods \cite{malladi2023mezo} use forward evaluations. Other methods use gradients to update only biases or adapters \cite{benzaken2022bitfit,hu2022lora}. Agents perform software engineering \cite{yang2024sweagent}, machine-learning engineering \cite{chan2024mlebench,jiang2025aide} and recommender development \cite{pan2026recevolve}. Here ES searches selected parameters, while weight coding uses model structure and feedback to propose, execute and refine edits.

\section{Methodology}\label{sec:3}
\subsection{Problem Formulation}\label{sec:3.1}

Let $f_\theta$ be a model with trained or randomly initialized parameters $\theta$. An edit $e: \theta \mapsto \theta'$ rewrites selected parameters while leaving the rest unchanged. In this study, model architecture and features remain fixed for simplicity and so does the serving logic and infra. For a production metric $M_t$ on day $t$, weight coding seeks $\Delta_t(e)=M_t(e(\theta))-M_t(\theta)>0$ on days excluded from edit selection. Our main retrieval metric is recall@$k$ micro-averaged over engagement types (Sec.~\ref{sec:4.1}) while in practice, other metrics are introduced for different models and business goals when necessary.

An edit's \emph{support} $S(e)$ contains changed parameter entries, so $|S(e)|$ counts changed values. Its \emph{degrees of freedom} count independently adjustable values or controls, given the selected sites and initialization. Resetting a 64-dimensional layer normalization (LayerNorm) gain \cite{ba2016layernorm} to one can change 64 values with no free parameter. Reinitializing a module can change millions of values while subsequent edits may only change a few thousand.

\subsection{Edit Primitives}\label{sec:3.2}

An edit primitive is a reusable weight transformation applied direct to the model. A concrete edit specifies both target parameters and numerical settings. Table~\ref{tab:1} gives important examples used in this work, with $w$ denoting the selected parameter block and $c$, $\lambda$ and $\alpha$ are scalar parameters to control the edits.

\begin{table}[htbp]
\caption{Important example edit operations.}
\label{tab:1}
\small
\begin{tabular}{@{}lp{0.73\columnwidth}@{}}
\toprule
Primitive & Example operation \\
\midrule
Rescale & $w \leftarrow c\,w$ \\
Reset & Replace $w$ with defaults or a fresh random initialization \\
Remove & Zero a selected branch's output projection \\
Compress & Low-rank truncation, pruning or quantization of $w$ \\
Interpolate & $w \leftarrow (1-\lambda)w+\lambda\bar w$, using compatible head or checkpoint weights $\bar w$ \\
Steer & $w \leftarrow w+\alpha u$ along a chosen direction $u$ \\
Search & Tune values or row scales by evaluation \\
\bottomrule
\end{tabular}
\end{table}

While most edits are model- and usecase-agnostic, their design and consequence can be tailored to specific application. In our example, while the two-tower retriever is only designed to multi-task on engagement types, steering the head's bias $w=b$ can push the retrieved candidates towards fresh or product Pins. Specifically, here steer uses a concept direction $u_F=(\mu_F-\mu_{\neg F})/\lVert\mu_F-\mu_{\neg F}\rVert$, where $\mu_F$ and $\mu_{\neg F}$ denote mean item embeddings with and without an attribute $F$ like freshness.

Primitive design determines which model changes can be expressed and searched. Agents can design and test new primitives tailored to model, infra and business goals. A agent could also revise its own edit-generation procedures and tools using evaluation feedback, then reuse validated improvements in later iterations toward recursive self-improvement (RSI) of the editing process \cite{zelikman2024stop}.

\subsection{Agent and Workflow}\label{sec:3.3}\label{sec:3.4}

Given goals and constraints stated by engineers in natural language, an LLM coding agent (Opus 5.5 with Max thinking) inspects model structure, proposes edits and uses evaluation feedback to refine them (Fig.~\ref{fig:1}(a)). It works through a sandboxed shell with controlled access to production code repository, training platform, query engine and model artifacts.

\textbf{Scan.} The agent analyzes weights and evaluates primitives on module groups individually using a development (dev) day. \textbf{Filter.} It re-measures large effects on separate days and discards edits whose effects reverse sign. \textbf{Edit or search.} Stable effects guide hypotheses and explicit modifications. When a hypothesis identifies a site without specifying values, the agent searches it. To compose edits, it can continue searching with earlier edits in place. \textbf{Validate and ship.} Engineers select candidates passing validation protocol in Sec.~\ref{sec:3.6} for online tests and decide what ships from those results. If no valid candidates are found, the agent can revisit past hypothesis and results to scan again. It's also possible to include online results to guide future iterations.

Before the first scan, the agent built the edit-evaluate engine and checked it against the production evaluator. It registers and deploys edited models after engineer approves. An end-to-end two-tower iteration takes less than 2 hours before finding a handful candidates. Across iterations, the agent evaluated over 2,000 edit configurations and about 190,000 search candidates to achieve the progress shown in Fig.~\ref{fig:1}(b).

\subsection{Edit-Evaluate Engine}\label{sec:3.5}

Given the large size of both model and data of production models and evaluators, weight coding maintains its own edit-evaluate engine for faster iteration.
The engine runs the production model once on the production evaluator's data and caches embeddings and necessary intermediate outputs. Edits rewrite loaded TorchScript parameters in GPU memory, recomputing only affected modules and their downstream computations.

Search directly optimizes the production metric through selected values or a multiplicative scale for each row of a weight matrix. An evolution strategy \cite{salimans2017es} samples 16 antithetic pairs of Gaussian perturbations per step, scaled by each tensor's root mean square ($\sigma=0.05$) or applied to per-row log-scales ($\sigma=0.1$). It converts the 32 objective scores to centered ranks and updates with Adam \cite{kingma2015adam} at learning rate 0.03 to 0.05. Dev-day positives are randomly split into disjoint halves and search uses one half. Every 10 steps, the engine evaluates on the other half and retain the best step. Target-model evaluations require forward passes only.

\subsection{Validation Protocol}\label{sec:3.6}
To mitigate overfitting or reward hacking, a validation protocol is carefully designed in the loop. We use only dev days to support scanning, edit and search. Then the filter days only discard single edits whose effects reverse sign on dev and filter days. Finally, we also reserve the held-out days for full evaluation and to guide online experiment decision. The held-out days never guide iteration.

Following model-editing criteria \cite{meng2022rome,yao2023editing}, we measure efficacy on dev data and generalization on held-out days and metrics and corpora outside the search objective. Specificity checks cover engagement types, fresh, non-fresh and product Pins, distinct top-10 Pins and popular-item share. The evaluator ranks a positive by counting strictly higher candidate scores. A diagnostic edit collapsing item embeddings inflated recall@10 from 0.626 to 0.639, so we also use pessimistic tie-breaking, placing tied candidates ahead of the positive. Finally, the production evaluator scores the edited artifact, which is compared with production tensor by tensor.

\section{Experiments}\label{sec:4}

\subsection{Experimental Setup}\label{sec:4.1}

\textbf{Model.} Our main case is the 147M-parameter Image Search two-tower retriever. The query tower $q_\theta(x)$ embeds the user, query Pin and context, and the item tower $p_\theta(y)$ embeds candidate Pin $y$. Candidates are ranked by cosine similarity $s_\theta(x,y)=\langle q_\theta(x),p_\theta(y)\rangle$. Both towers start from a cross-surface foundation model and are fine-tuned on Image Search logs (Fig.~\ref{fig:2}(a)). The query tower processes 35 tokens through a two-layer DHEN (Fig.~\ref{fig:2}(b)). After an input LayerNorm, layer 0 (L0) combines attention and multilayer perceptron (MLP) branches with a shortcut that mixes all tokens. Layer 1 (L1) combines MLP and MaskNet branches with an identity shortcut. Each layer sums its branches and shortcut, then applies LayerNorm. A final MLP produces three L2-normalized output heads with a shared bias.

\textbf{Training.} Post-training finetune runs over 320M engagement logs with in-batch sampled softmax, all-gather negatives and logQ correction \cite{yi2019sampling}. Matrices use a Muon-family optimizer \cite{liu2025muon} while one-dimensional parameters, including LayerNorm parameters and learned temperatures, use Adam with decoupled weight decay \cite{loshchilov2019adamw}.

The cross-surface base model needs more than 200 GPU hours and a surface-level post-training run takes about 140 GPU hours across 8 A100. In comparison, a full loop spends 8.6 GPU hours on average and it took 4 loops to find the best model reported in this work, only 24.6\% of the finetune.

\textbf{Evaluation.} The production offline evaluator uses daily logs from both ranked and randomized traffic. It computes recall@$k$ within a fixed corpus for each engagement type and runs weighted micro-averages over engagement types. Gains are absolute recall changes in percentage points relative to production on the same evaluation dataset and configurations.

\begin{figure*}[t]
\centering
\includegraphics[width=\textwidth]{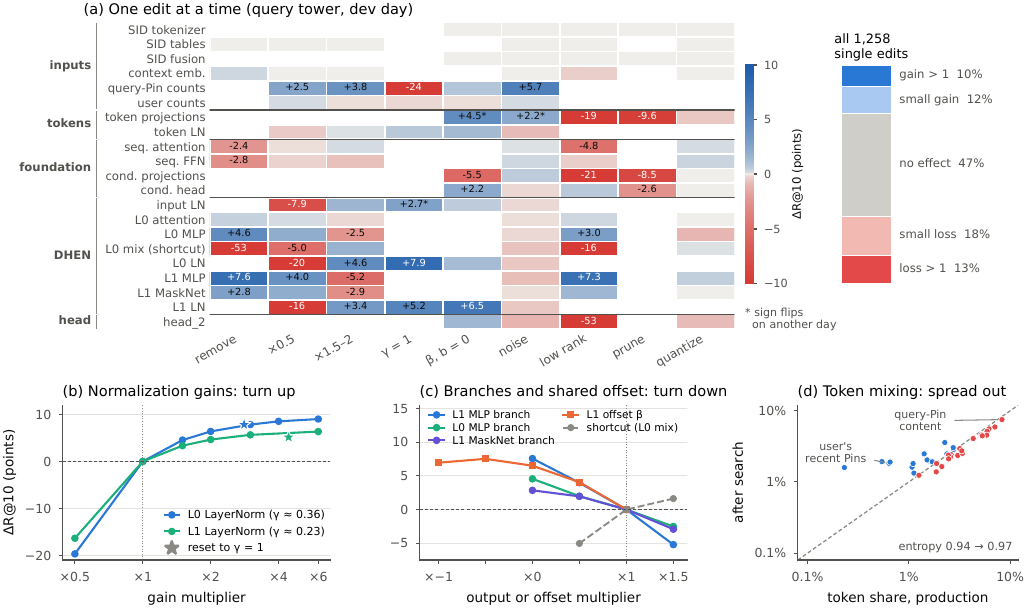}
\caption{Effects of direct edits. (a) Dev-day query-tower effects and the distribution of 1,258 edits across both towers. (b) LayerNorm gain scaling. (c) Branch, offset and shortcut scaling. (d) L0 shortcut token shares before and after mixing search.}
\Description{Four panels show single-edit effects, normalization gain scaling, branch and offset scaling, and token shares before and after search.}
\label{fig:3}
\end{figure*}

\subsection{Locating Editable Sites}\label{sec:4.2}

The scan covers 1,258 edits across 130 module groups (Fig.~\ref{fig:3}(a)). Of these, 47\% change recall@10 by less than 0.1 points, and only 10\% gain more than one point. Useful sites concentrate in DHEN normalization, branches and token mixing, and in the output head. The scan therefore narrows a large parameter space to structures whose contributions can be changed directly.

Filtering distinguishes persistent effects from gains specific to the dev sample. Of the re-measured edits whose absolute dev-day changes exceeded one point, 23\% reverse sign on another day. We filter these effects before developing or composing edits.

Past experience and editing results of other models can be used to guide the scan though the transfer can be negative. In practice, we find that allowing agent to access and analyze experiment results and source code often help.

\subsection{Editing Trained Weights}\label{sec:4.3}

\begin{table}[t]
\caption{Representative two-tower edits. r4 modifies the retrieval head, F04 resets normalization, L01 searches trained mixing weights, L07 re-searches L01 with r4, and L05 rebuilds randomized DHEN.
}
\label{tab:2}
\small
\centering
\begin{tabular}{lrrr}
\toprule
Variant & Changed & Searched & $\Delta$R@10 \\
\midrule
r4 & 65.6k & 0 & +8.83 \\
F04 & 128 & 0 & +10.10 \\
L01 & 2,064 & 2,064 & +13.69 \\
\textbf{L07} & \textbf{67.7k} & \textbf{2,128} & \textbf{+15.00} \\
L05 & 13.6M & 2,448 & +9.80 \\
\bottomrule
\end{tabular}
\end{table}

\textbf{Normalization and offsets.} DHEN LayerNorm gains in both L0 and L1 are below their initial value when the training is completed. Increasing them improves recall, while reducing the L1 offset also helps (Fig.~\ref{fig:3}(b,c)). These changes strengthen shortcut-carried tokens relative to subsequent branches and shared offsets. Shared offsets pull query embeddings toward a common direction and can concentrate retrieval on hub Pins \cite{conneau2018word}. Edit F04 resets the L0 gain and the L1 offset, and changing all these values gains 10.1 points under the production evaluator (Tab.~\ref{tab:2}).

The head edit r4 changes a related shared component. It shifts the retrieval head's bias away from the mean embedding of non-fresh Pins and toward the fresh direction, adds a constant shift, and mixes 20\% of the weights of another trained output head. Its four scalar settings yield a gain of 8.8 points.

\textbf{Branches.} Reducing branch contributions and increasing the L0 shortcut contribution improve recall (Fig.~\ref{fig:3}(c)). Removing all four branches leaves linear token mixing and LayerNorms and gains 6.1 points across all held-out days, with every day above production. Thus, the trained branches make a negative net contribution under this metric. Edit F07 combines low-rank truncation and pruning of the L1 MLP branch with edit r4. Edit L06 additionally searches the three LayerNorms. These edits establish that direct changes to a few computational components can improve a trained model.

\subsection{Searching and Composition}\label{sec:4.4}

\subsubsection{Searching Trained Weights} Token-mixing weights concentrate on a few inputs, motivating a search beyond the discrete edits. Searching five DHEN mixing matrices of 2,064 values, from trained weights for 150 steps gives edit L01, with a production gain of 13.7 points. The matrices change by 39\% to 64\% in relative norm, while their root mean square changes by only $-$6\% to +16\%. Thus, the search mainly redistributes contributions across tokens. In the L0 shortcut, a token's share is its fraction of total absolute mixing weight. Smaller shares increase and dominant shares decrease, raising normalized entropy from 0.94 to 0.97 (Fig.~\ref{fig:3}(d)). This changes how learned features contribute without relearning their representations from scratch.

\begin{figure}[t]
\centering
\includegraphics[width=\columnwidth]{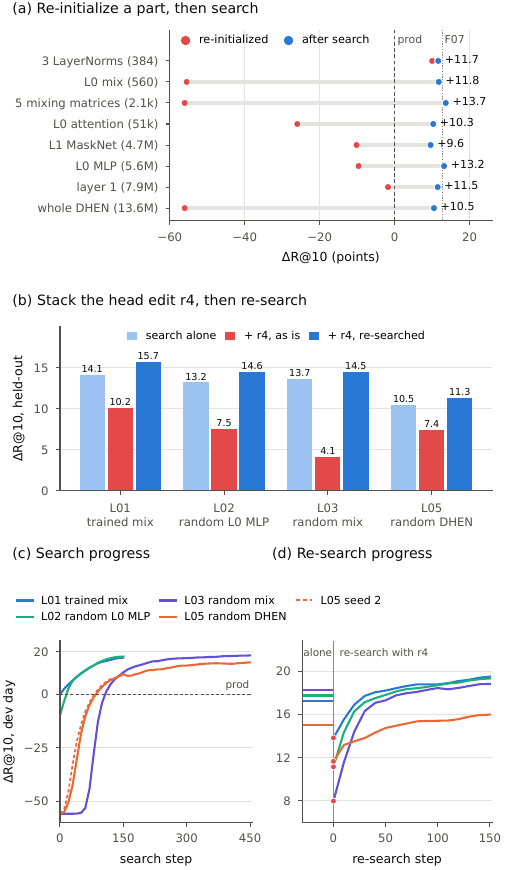}
\caption{Search and composition. (a) Recall after re-initialization (red, dev day) and search (blue, held-out days). (b) Adding r4 before and after re-search. (c) Search trajectories from trained and random starts. (d) Re-search after adding r4.}
\Description{Four panels compare recovery after random initialization, composition before and after re-search, search trajectories, and re-search trajectories.}
\label{fig:4}
\end{figure}

\subsubsection{Rebuilding Randomized Modules}\label{sec:4.4.random}

We next test whether direct editing can rebuild effective computation from random weights. We re-initialize a module with the default PyTorch initializers while retaining the surrounding trained model. For certain module, re-initializing it effectively reduces dev-day recall to zero while others have less impact (Fig.~\ref{fig:4}(a)).

Edit L03 searches 2,064 randomly initialized mixing values. L02 re-initializes the 5.6M-parameter L0 MLP branch and searches 12,688 row scales. The choice of editable values matters. Searching only row scales of the whole random DHEN remains below production at -10.0 points as row scaling cannot change how tokens are mixed. Successful mixing matrices have correlations of 0.13 to 0.46 with the trained ones, while their changes in token shares correlate at 0.76 with those found from trained starts. Different matrices can therefore provide useful mixing through similar redistributions of token contributions.

After reinitializing all 13.6M DHEN parameters, searching only its mixing matrices and LayerNorms with 2,448 values, gives edit  L05. It gains 9.8 points under the production evaluator and 10.5 over held-out days. Most parameters in the rebuilt interaction module remain random, while its trained encoders, output head and item tower are retained. This result extends weight coding from improving trained parameter values to constructing useful interaction computation within an inherited model.

Fig.~\ref{fig:4}(c) shows recovery from random starts. Mixing and whole-DHEN searches initially remain near zero recall, then rise sharply and cross production after 90 to 110 steps. The two whole-DHEN search with different seeds follow similar trajectories.

\subsubsection{Composition of Edits} Combining edit r4 directly with L01 lowers its four-day gain from 14.1 to 10.2 points (Fig.~\ref{fig:4}(b)). Independently developed edits can make overlapping corrections to shared offsets and the relative strength of shortcuts. With r4 in place, we re-search the mixing matrices together with the retrieval head's bias following the same setup. The resulting edit L07 gains 15.0 points under the production evaluator and 15.7 over held-out days, with +13.8 on the worst day. Re-search lets the edited components adapt to one another while simple stacking doesn't always work.

The trajectories in Fig.~\ref{fig:4}(d) characterize compositions across all four starts, where similar evolving trends are observed, validating the search process in the large weight space.

\subsection{Across the Recommendation Funnel}\label{sec:4.5}

\begin{table}[t]
\caption{Weight coding gains for generative retrieval PinRec, and for the Image Search ranker on hold-out days.}
\label{tab:3}
\small
\centering
\begin{tabular}{llr}
\toprule
Model & Metric & Change \\
\midrule
PinRec & Unordered R@1 & +2.31 pp \\
       & Ordered R@1 & +2.64 pp \\
Ranker & Utility NDCG & +0.53\% \\
\bottomrule
\end{tabular}
\end{table}

Two further production models test the process across the funnel, using their own evaluators and the same date split. PinRec provides generative retrieval, while the ranker scores candidates from all retrieval sources and rank them using a blending model.

\textbf{PinRec.} This 106M-parameter generative retriever \cite{botta2025pinrec} uses an autoregressive transformer and a shared Pin embedder. Unordered recall@$k$ counts a target if any query embedding ranks it in the top $k$, while ordered recall matches each query to the target at its autoregressive step.

\begin{figure}[t]
\centering
\includegraphics[width=\columnwidth]{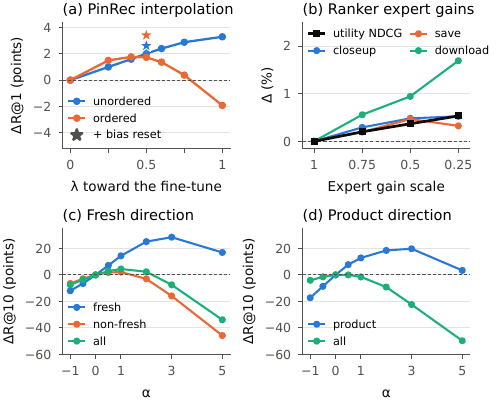}
\caption{Edits across models and objectives. (a) PinRec interpolation from production ($\lambda=0$) to its fine-tune ($\lambda=1$). (b) Relative changes in ranker utility NDCG and hit@3 after expert scaling and recalibration. (c,d) Two-tower recall@10 under fresh and product steering.}
\Description{Four panels show PinRec interpolation, ranker expert scaling, and recall under freshness and product steering.}
\label{fig:5}
\end{figure}

The edits must reflect this architecture. Resetting the Pin embedder's LayerNorm gains costs 8.8 points of unordered recall@10 on the dev day. Weight inspection reveals biases excluded from weight decay and late matrix updates below bfloat16 resolution \cite{kalamkar2019bf16}. A related Image Search fine-tune used for ads provides another snapshot. Zeroing the output bias and interpolating halfway toward that snapshot \cite{wortsman2022soups,wortsman2022wiseft} each improve every metric on every held-out day. The interpolation curve shows why the intermediate snapshot matters (Fig.~\ref{fig:5}(a)). Unordered recall@1 improves toward the fine-tune, but ordered recall@1 peaks before it and eventually falls below production. Combining halfway interpolation with a reset of the output head's hidden-layer bias improves both held-out R@1 metrics (Tab.~\ref{tab:3}).

\textbf{Ranker.} Ranker is the largest model we tested with 1.37B parameters and it's designed to predict the engagement probably of multiple events. Its multi-gate mixture of experts \cite{ma2018mmoe} has eight experts and 15 engagement heads. Each head combines a gated expert mixture with a direct path, followed by calibration and a fixed utility blend. Calibration alone raises a binned estimate of the AUC by up to 0.11 in absolute units on rare heads without changing rankings. Small edits also change top-3 metrics on only about 6\% of requests. We therefore use exact request-level AUC and normalized discounted cumulative gain (NDCG) with utility gains, refitting each head's calibration on the dev day before comparing blends.

Resetting expert LayerNorm gains to one reduces closeup AUC by 0.029 in absolute units. Reducing them weakens experts relative to the direct path and improves ranking (Fig.~\ref{fig:5}(b)). The half- and quarter-scale edits improve utility NDCG on every held-out day (Tab.~\ref{tab:3}), with corresponding $z$-scores ranging from 7.8 to 10.5. Hit@3, the number of engaged Pins among each request's top three candidates per user, also increases for closeups, saves, downloads and shares.

These results show that weight coding adapts to different architectures and metrics, with distinct changes improving generative retrieval and final ranking. On the other hand, many useful edits on two-tower retriever do not generalize to PinRec and the ranker.

\subsection{Steering Edits}\label{sec:4.6}

We now turn to look at how steer edits could provide more precise control across segments. A scalar $\alpha$ can control each direction for freshness or shoppable products, and we evaluate the full grid on held-out days. Offline exposure is the share of retrieved top-10 Pins from the given segment. These edits provide a new route to adjust exposure targets directly, comparing to budget and triggering optimization at pre-retrieval stage \cite{hou2026causal}.

At $\alpha=1$, freshness steering improves overall recall@10 by 4.5 points and fresh recall by 14.4, while fresh exposure rises from 5.2\% to 13.4\%. Product steering improves product recall by 12.8 points at a cost of 1.5 points overall, with product exposure increasing from 0.26\% to 1.06\%. Fig.~\ref{fig:5}(c,d) shows how larger shifts change these trade-offs.

Exposure changes monotonically over the tested $\alpha$ range. Fresh share in the top-10 spans 2\% at $\alpha=-1$ to 70\% at $\alpha=5$, while product share reaches 39\% at $\alpha=5$. Strong product steering also concentrates results. The share of the top-1\% most popular Pins rises from 0.10 to 0.42 at $\alpha=5$, delimiting the useful range of this control.

Steering also composes with searched edits. Adding the product direction at $\alpha=1$ to L07 raises its product recall gain from +28.3 to +38.2 points with an overall gain of +14.1. Product recall estimates are less precise than exposure measurements because this evaluation has fewer product positives per day.

\subsection{Online Results}\label{sec:4.8}

\begin{table}[t]
\caption{Online A/B changes relative to production for the main two-tower source. Counts are per user, while product and fresh shares are fractions of source impressions. Fresh Pins are at most 28 days old (1\% of users).}
\label{tab:6}
\small
\setlength{\tabcolsep}{3pt}
\centering
\begin{tabular}{@{}lrrrrrr@{}}
\toprule
Arm & Impressions & Closeups & Saves & \shortstack{Long\\clicks} & \shortstack{Product\\share} & \shortstack{Fresh\\share} \\
\midrule
\multicolumn{7}{l}{\emph{Recall over all engagement types with a focus on closeup}} \\
r4 & $-$2.0\% & +0.9\% & $-$3.5\% & +68.0\% & +91.3\% & +28.9\% \\
r6 & +0.4\% & +3.2\% & $-$0.9\% & +54.2\% & +63.6\% & +14.6\% \\
F04 & $-$1.9\% & +1.8\% & $-$3.9\% & +37.2\% & +34.4\% & $-$8.5\% \\
L01 & +0.8\% & +5.6\% & +1.0\% & +69.9\% & +71.3\% & $-$12.3\% \\
L06 & $-$1.3\% & +3.9\% & $-$1.2\% & +94.4\% & +158.5\% & +8.5\% \\
L07 & +0.2\% & +6.7\% & $-$0.3\% & +97.0\% & +147.9\% & $-$1.6\% \\
\midrule
\multicolumn{7}{l}{\emph{Save-type recall with capped clicks, long clicks and products}} \\
A2 & +5.2\% & +7.6\% & +10.7\% & $-$24.9\% & $-$37.8\% & +8.6\% \\
B2 & +4.0\% & +4.8\% & +8.4\% & $-$14.6\% & $-$32.5\% & $-$1.0\% \\
C2 & +4.7\% & +6.6\% & +9.6\% & $-$23.2\% & $-$39.8\% & +0.1\% \\
\bottomrule
\end{tabular}
\end{table}

We bring the two-tower model to online A/B test with real production traffic and Tab.~\ref{tab:6} reports attributed engagement, per user and relative to production. The results of each group are measured on around 10M users for about a week.

\textbf{All-engagement objective.} Six arms use edits selected to improve recall over all engagement types, including r6, a weaker version of r4. Due to the data distribution, those models are optimized more towards Closeup and the online results show that every arm raises closeups per impression. Across arms, this rate tracks the production recall@10 gain with Pearson correlation 0.97. The edits also increase long clicks and product share, while saves can fall. Fresh share rises for r4, r6 and L06, which carry the fresh direction of r4, falls for F04 and L01, and stays close to neutral for L07, which combines the direction with the mixing search of L01.

\textbf{Save-oriented objective.} To target saves and organic content, we change the evaluation objective and use positives from production traffic with the serving index of more than 50M Pins. The agent searches the same 2,128 values as L07 including the five mixing matrices and the retrieval head's bias. The target is recall of the evaluator's save-type engagement label, with hinge penalties on clicks, long clicks and products. Three searches with different starts and corpus weights produce A2, B2 and C2.

These edits raise saves from the source by 8.4-10.7\% per user and 4.3-5.3\% per impression. They also increase closeups while reducing long clicks and product share. The same editable sites therefore support different business goals under different evaluation objectives. These outcomes extend weight coding from improving model computation to aligning its preference through evaluation-guided edits, without a full post-training.

One caveat to call out is that the ranker tends to rank the Pins from the edited model much higher and such a cannibalization reduces impression share and engagements from other retrievers. For downstream models like pre-ranking and ranking, we don't need to worry about this. Our online experiments of weight-coded pre-ranking models show strong site-wide gains.

\section{Discussion}\label{sec:5}

% Weight coding connects improvements at a few trained sites, composition of edits and useful computation rebuilt within a randomized module. These results suggest a broader role for direct edits in model development, with distinct questions about generalization, mechanism, production use and future research questions.

\subsection{Generalization and Reward Hacking}\label{sec:5.validation}

Metric optimization can exploit evaluation artifacts or improve a proxy without improving the intended behavior \cite{gao2023overoptimization,manheim2018goodhart,pan2026recevolve}, which could lead to degraded online results and bring bad user experience. To test if weight coding exploits reward hacking, we run evaluation on weight-coded models using a spectrum of metrics, including those are not used in the process, on unseen dataset and corpus. In the offline two-tower validation, most tested variants improve all ten engagement types on all held-out days, and both fresh and non-fresh engagement improve every day. The gain of searched edits' recall@10 falls by 4.5 points on average from the dev day to held-out days but still remains positive across held-out days.

We further prepare separate corpus for evaluation and the recall@10 gains persist, reaching +9.7 points for L07. Gains also hold on the full corpus and on recall@1000, neither of which was used or optimized during weight coding process. Distinct Pins in the top-10 increase by 9-18\%, while the share of the top-1\% most popular Pins falls. In conventional training iteration, it's uncommon to see such large gains across overall recall, diversity and freshness.

While the gains across dates, corpora and metrics support generalization within the tested setting, it doesn't rule out adaptive overfitting or biases shared across dates. In practice, we find that such a cross validation is enough to select candidates for online experiments while early retriever models can still suffer from severer online-offline consistency. 

\subsection{Implications for Model Development}\label{sec:5.1}

The random-DHEN result provides an interesting perspective. Its encoders, output head and item index remain fixed, while searched mixing and normalization parameters establish useful interactions. Together with different edits across the recommendation funnel and online goal adaptation, this supports direct editing as a way to develop model and control behavior.

For trained weights, we conjecture that a single training schedule can leave normalization gains, shared offsets and branch strengths poorly balanced for the target task. Edits adjust their relative contributions while retaining representations learned elsewhere. Resetting the L0 gain (edit F02) lowers the in-batch sampled-softmax loss of the retrieval head on the last training day from 1.300 to 1.015. Normalization and branch edits also give similar recall gains on training days and a held-out day a month later.

A retrain without weight decay on one-dimensional parameters has LayerNorm gains near one but loses 3.5 recall@10 points under the production evaluator. Mixing-matrix search still adds 16.8 points, compared with 14.1 on production. Small normalization gains alone therefore do not explain the room for weight coding. PinRec benefits from bias edits and interpolation, while resetting the ranker's small LayerNorm gains hurts. Successful edits on the ranker weaken experts relative to their direct path. As a result, useful edits tend to be model-specific while the weight-coding methodology applies generally. It remains an important question to understand how generally weight coding applies to different models, tasks and datasets. Taking a step further, how it complement or even replace the full training, either pre- or post-training. 

\subsection{Productionization and Future Directions}\label{sec:5.3}
While this work establishes weight coding in real production setting to evolve large-scale models end-to-end with online validation, there are still critical steps to follow before it can become a generally useful methodology in engineering practice.

Establishing it as routine production practice requires understanding which modules admit effective control through few parameters and whether successful edits remain useful after model / checkpoint refreshes or data / user distribution shifts. This would help turn the observed progression from local edits to module reconstruction into a more predictable development strategy.

Edits need a reproducible lifecycle tied to their starting checkpoint, code, evaluation data, objective and verified artifact. Reusing an edit requires revalidation and attention to dependent artifacts, as the PinRec and ranker examples illustrate. Monitoring and rollback procedures should determine when to retain, revise or retire edits as the model and ecosystem change over time.

Joint weight coding across corpus, indexing, retrieval model and ranker could help synchronize the funnel, coding both architecture and weights opens routes for new models (in one test, we reduce the model size of the two tower by 65\% without hitting offline performance), and searching over multiple objectives can help align better with business goals and long-term use value. While online feedback are useful in this work, it was not incorporated automatically in the loop. RSI can be introduced to weight coding to explore richer edits and build tools and systems to continually improve by itself. Our example on randomized interaction module opens the possibility of weight-coding a full model from scratch (all random weights) and the scaling law behind it remains a mystery. But finally, important launch, product and business decisions should still be reviewed by human engineers and experts.

\section{Conclusion}\label{sec:6}

Weight coding develops models through executable weight edits guided by an LLM agent and evaluation feedback. On production recommenders, it improves trained weights, constructs effective interaction mappings after random re-initialization, and adapts edits across retrieval and ranking. Steering and online results establish control over exposure and gains aligned with business objectives, supporting a novel path for efficient model evolution.\label{paper:main-end}

\section{Acknowledgement}
J. Hou would like to extend special thanks to Xiao Sai, Roberto Konow and Yi-Chin Wu for great support and comment of this work, and Olafur Gudmundsson, James Ding and Xianxing Zhang for insightful discussions.

\bibliographystyle{ACM-Reference-Format}
\bibliography{refs}

\end{document}